# Reversible Sulfuric Acid Doping of Graphene Probed by in-situ Multi-Wavelength Raman Spectroscopy

Gwanghyun Ahn[1] and Sunmin Ryu[1,2*]

[1]Department of Chemistry, Pohang University of Science and Technology (POSTECH), Pohang, Gyeongbuk 37673, Korea

[2]Division of Advanced Materials Science, Pohang University of Science and Technology (POSTECH), Pohang, Gyeongbuk 37673, Korea

**Abstract**

Since lattice strain and charge density affect various material properties of graphene, a reliable and efficient method is required for quantification of the two variables. While Raman spectroscopy is sensitive and non-destructive, its validity towards precise quantification of chemical charge doping has not been tested. In this work, we quantified in-situ the fractional frequency change of 2D and G peaks in response of charge density induced by sulfuric acid solution as well as native lattice strain. Based on the experimental data and theoretical corroboration, we presented an optical method that simultaneously determines strain and chemically-induced charge density for three popular excitation wavelengths of 457, 514 and 633 nm. In order to expedite intercalation of dopant species through the graphene-SiO$_2$ substrates, dense arrays of nanopores were precisely generated in graphene by thermal oxidation. The nano-perforated graphene membrane system was robust for multiple cycles of doping and undoping processes, and will be useful in studying various types of chemical interactions with graphene.

## 1. Introduction

Because of low density of electronic states near its Fermi level, various material properties of graphene is severely affected by excess charge carriers.[1] Manipulation of charge density in graphene materials has been studied to improve electrical conductivity,[2] modulate optical transmission,[3] reveal otherwise unoccupied bands by photoemission,[4] and introduce asymmetry and thus open bandgaps in bilayers.[5] In particular, electrostatic charge doping in a parallel capacitor geometry[6] has been exploited in early studies for bandgap manipulation[7] and Raman scattering[8] for single and few-layer graphene. Because of its electrical circuitry, the method allows charge density to be varied reversibly and instantaneously towards either of the two polarities in a quantitative manner. Using electrolytes of high effective dielectric constant, the charge density can reach $\sim 10^{14}$ /cm$^2$,[9] which enabled observation of structural[10] and superconducting[9] phase transitions of layered crystals. As an alternative, charge transfer doping using various electron-donating or -withdrawing chemicals have been tested and mainly investigated for better electrical conduction of graphene as a transparent electrode.[11] Because of the charge transfer binding between dopants and graphene, the chemical complex is somewhat stable in the ambient conditions without further treatments, which can be very important in applications requiring long-lasting effects.[2] Typical chemical treatments such as vapor deposition, dipping in solutions and coating with polymers can be readily applied to large-area samples and scalable for samples of large quantity as an industrial process. In addition, there are many n- and p-type chemical dopants in gaseous, liquid and solid forms[11] so that appropriate candidates may be selected for applications with chemical compatibility issues.

The density of extra charges ($n$) in graphene can be determined either by electrical[12] or optical[8] methods. The former uses minimum conductivity at Dirac point or Hall effect in the field-effect transistor or Hall-bar geometries, respectively. Raman spectroscopy, one among the latter, has obvious strengths over the electrical methods in the sense that it does not require prior treatments or fabrication steps, and thus can probe samples in virtually arbitrary forms. The method can also provide a high spatial resolution only limited by the optical diffraction limit and be used in gaining spatial distribution of charges. In particular, the multimodal sensitivity issue of Raman G and 2D frequencies ($\omega_G$ & $\omega_{2D}$), towards charge density[8, 13] and lattice strain[14] ($\varepsilon$), has been resolved via a vector analysis under the assumption of mutual independency between the two.[15] Since the analysis framework is based on the Raman data from electrically-tuned graphene,[16] however, its validity needs to be tested against chemical modulation of charges. While some previous studies looked at the relation between charge doping and Raman spectral shift,[17-21] they failed to fulfill an important requirement for the test, spatially-resolved in-situ measurements during doping and undoping processes, which can then provide reliable calibration data. This condition is necessary since typical graphene samples exhibit sizable spatial inhomogeneity in native strain[15] and charge density.[22] In addition, most reference Raman data that were obtained with green excitation wavelengths, 514 or 532 nm, cannot be used in quantification of charge density and strain with other wavelengths because of the large frequency dispersion of 2D peak.[23]

In this work, we developed a simple model system for reversible chemical charge transfer using nano-perforated graphene supported on substrates. In-situ Raman spectroscopic measurements allowed us to monitor p-type doping of graphene by $H_2SO_4$ and its reversal in real-time manner. Our results showed that hole doping occurs at both surfaces of graphene by sulfuric acids that diffuse through nanopores and is highly reversible. The fractional frequency variation, $(\Delta\omega_{2D}/\Delta\omega_G)_n$, was found to be virtually constant as $0.70 \pm 0.06$ over the entire region of *n*, which agrees well with theoretic prediction and previous results obtained by the electrical method.[16] We also established independent sets of metrology data obtained at three popular Raman excitation wavelength, 457, 514 and 633 nm, which can be applied to an arbitrary wavelength.

## 2. Methods

Graphene samples were prepared by mechanically exfoliating kish graphite on to Si substrates terminated with 285 nm-thick $SiO_2$ layer.[24] The number of layers and crystalline quality of graphene were determined by Raman spectroscopy.[25] The microscope-based setup has been described in detail elsewhere.[26] Briefly, excitation laser beam was guided onto samples with a focal spot of ~1 micron by an objective (40X, numerical aperture = 0.60). Three different wavelengths were used: 457 nm from a diode-pumped solid state laser, 514 nm from an Ar ion laser and 633 nm from a He-Ne laser. Scattered signals were collected by the same objective and guided to a spectrograph (focal length = 300 mm) equipped with a charge-coupled device. The linewidth of the Rayleigh peak was 3.0 cm$^{-1}$ for the 514 nm laser and the overall spectral accuracy of the measurements was better than 0.5 cm$^{-1}$. The average power on samples was maintained below 1.7 mW, which generated no detectible photo-induced effects. Unless stated otherwise, presented data were obtained with the 514 nm laser. Raman peaks were fitted with single or double Lorentzian functions depending on the line shapes.

To generate nanopores in graphene membranes, samples were partially oxidized at 550 °C in flowing $Ar:O_2$ mixture gas (600:150 mL/min) according to the previous work.[27] Atomic force microscopy (AFM) revealed topographic details of nano-perforated graphene with randomly distributed holes of ~100 nm in diameter and ~30 $\mu m^{-2}$ in number density. To form an optical liquid cell (see Fig. S1) with two entrances that allow variation of concentration, substrates with suitable graphene samples were placed upside down on top of a slide glass with glass spacers of 200 μm in thickness as described elsewhere.[27] Deionized (DI) water or $H_2SO_4$ solution of various concentration was introduced through the edges of the cell into the 200-μm-high space in dropwise manner. To increase (decrease) its concentration, concentrated solution (DI water) was added through one entrance while withdrawing the liquid in the cell through the other entrance with absorbing filtering paper.

## 3. Results and Discussion

Figure 1a and 1b show an optical micrograph of a typical 1L graphene sample and Raman spectra of 1L in sulfuric acid of various concentration ($c_S$). Pristine graphene exhibited prominent G and 2D peaks that originate from zone-center $E_{2g}$ mode and overtone of transverse-optical mode near K points, respectively.[25] High 2D/G intensity ratio and other spectral features indicated that the charge density is very low (~$1 \times 10^{12}$ /$cm^2$) as will be explained below. In addition, defect-activated D peak was negligible, which indicates its high structural quality. When immersed in DI water, there was no significant spectral change. In 1 M solution of $H_2SO_4$, G peak upshifted and became narrower in its line shape. At the same time, the 2D/G intensity ratio decreased significantly. All these changes are consistent with increased hole density due to charge transfer doping in sulfuric acid. As increasing the concentration of the acid up to 9 M, the two changes became more obvious suggesting further increased hole density. For 3 M acid, a rather broad peak emerged near 1040 $cm^{-1}$, which originates from $HSO_4^-$ ions.[28] As shown in Fig. S2, the Raman peaks for water centered at 3300 $cm^{-1}$ also changed their line shape significantly at high concentrations. Because of the finite focal depth of the objective (~1 μm), the majority of the Raman signals from bisulfate ions and water are not related to graphene, but serve as a local indicator for the solvent environment.

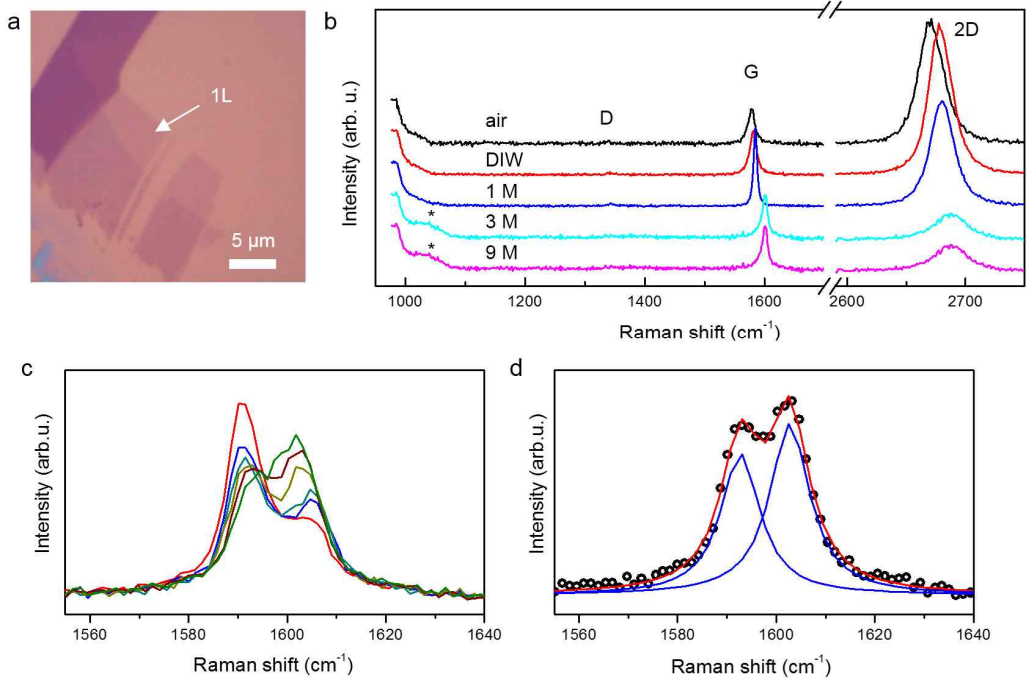

**Figure 1**. Surface vs intercalative charge transfer doping processes with distinctive kinetics. (a) Optical micrograph of a typical graphene (1L) sample. (b) G and 2D Raman spectra of pristine graphene in the ambient air, deionized water (DIW) and $H_2SO_4$ solution of various nominal concentration. The asterisk-marked peak at ~1040 cm$^{-1}$ originates from bisulfate ions ($HSO_4^-$). (c) 10 time-coursed G band Raman spectra obtained for 600 s right after injecting 9 M solution. (d) The fifth spectrum in (c) was nicely fitted with two Lorentzian components.

We found that the dopants intercalate through the graphene/$SiO_2$ interface and add inhomogeneity to charge transfer kinetics. Figure 1c show time-course spectra of G peak obtained after 2 drops of 9 M $H_2SO_4$ solution were added to the optical cell. Over 10 minutes, another G peak appeared at a higher frequency ($\omega_G$) of 1603 cm$^{-1}$ instead of gradual upshift of the peak centered at 1593 cm$^{-1}$. Notably, the fifth spectrum shows roughly 1:1 contribution from the two Lorentzian components as decomposed in Fig. 1d. The last spectrum is still asymmetric and consists of the two sub-peaks. This indicates that charge density is inhomogeneous across the probed graphene area and presumably bimodal. The delayed additional doping represented by the high-frequency G peak in Fig. 1c can be explained by interfacial diffusion of the dopants.[19] The top surface of graphene supported on $SiO_2$ substrate is freely accessible to molecular species in the solution. However, the other surface facing the substrate is completely sealed off except the interfacial channels. Recent studies showed that the interfacial space with a gap distance of a few angstroms may accommodate some molecular intercalants such as water,[27] halogens[29] and oxygen.[15] Zhao et al. also attributed asymmetric doping of 2L graphene by sulfuric acid to the possibility that different kinetics apply to both surfaces.[19] Since bisulfate ions and $H_2SO_4$ molecules co-intercalate in 1:2.5 number ratio inducing hole carriers in graphite,[30] similar doping mechanism may be operative. But detailed mechanistic understanding requires systematic

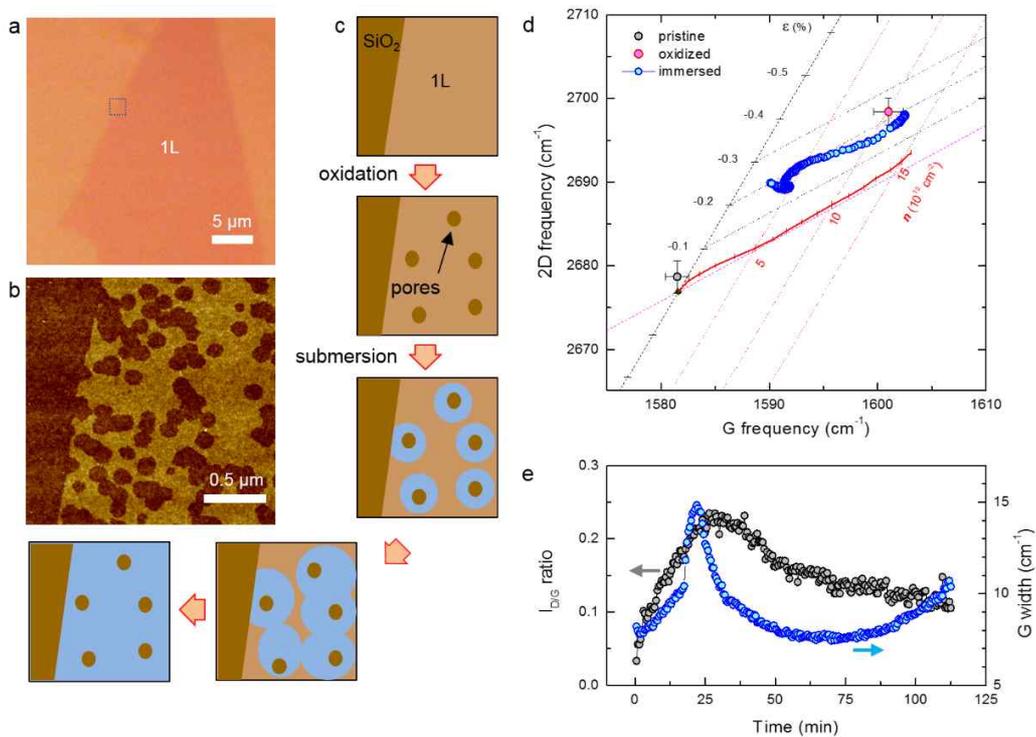

**Figure 2.** Intercalation of water accelerated in nano-perforated graphene. (a) Optical micrograph of perforated graphene. (b) Structure of nano-perforated graphene. The AFM height image was obtained from a graphene sample oxidized at 550 °C for 30 min. (c) Schematic diagram of water intercalation through the nanopores. (d) Time-coursed measurements of ($\omega_G$, $\omega_{2D}$) for ~2 hrs (30 s for each measurement) after immersion in DI water. Data for pristine and oxidized states obtained in the ambient conditions were included with statistical errors obtained from ~10 μm² area. (e) Intensity ratio ($I_D/I_G$) of D and G peaks (black circles), and linewidth of G peak (blue circles) obtained as a function of the immersion time.

investigation, which goes beyond the scope of this work.

In order to achieve faster and spatially homogeneous response to concentration change, graphene in Fig. 2a was perforated by controlled thermal oxidation at 550 °C (Fig. 2b).[27] As depicted in Fig. 2c, numerous nanopores of ~100 nm in diameter were created and served as an entrance to expedite interfacial diffusion of the dopants. Since the average distance between neighboring nanopores was shorter than a few hundred nm, molecular species can readily access the interfacial space between the graphene membrane and substrate through the nanopores. On the other hand, its Raman spectra showed noticeable changes with respect to the pristine state as shown in Fig. S3. Most of all, D peak was activated by the edges of nanopores throughout the whole sample area (see Fig. S4 for Raman map of D/G area ratio). G peak upshifted and became very narrow. 2D peak upshifted even more and decreased in intensity drastically. All these changes except the D peak can be explained by oxygen-related charge transfer reaction that is induced by thermal activation (see Raman maps of $\omega_G$ and $\omega_{2D}$ for their spatial distributions in Fig. S4.).[31] Figure 2d showed that the hole density of oxidized graphene amounts to $\sim 1 \times 10^{13}$ cm$^{-2}$, which may interfere with targeted charge exchange between graphene and sulfuric acid. To remove the thermally-induced charges, the sample was placed in DI water using the optical cell described in the Methods. According to a previous study,[27] the thermally activated charge transfer can be reversed by intercalation of water that drives out the hole dopants including oxygen. While its mechanistic details have yet to be revealed, our data confirmed the reversal that was initiated by immersion in water: the real-time trajectory of ($\omega_G$, $\omega_{2D}$) in Fig. 2d was aligned in parallel with the *n*-axis representing the density of hole carriers,[15] and hole density decreased to $\sim 4 \times 10^{12}$ cm$^{-2}$ during the first 2 hours in water. The notable winding in the trajectory is related to lattice distortion induced by interfacial diffusion of water.[27] Since graphene supported on SiO$_2$ substrates are highly flat with a typical root-mean-squared roughness of 0.15 nm,[32] a single water molecule with a van der Waals diameter of ~0.3 nm[33] will impose non-negligible structural perturbation when situated in the interfacial space. The gradual increase and recovery of D/G Raman intensity ratio in Fig. 2e illustrates the distortion and relaxation of the graphene lattice during the course of water intercalation.[27] The linewidth of the G peak reached a maximum at ~22 min because of increasing spatial heterogeneity in charge density and lattice strain, and its gradual increase after 80 min is due to the intrinsic line broadening that is characteristic of charge-neutral graphene.[8]

Using nano-perforated graphene samples, we monitored the trajectory of ($\omega_G$, $\omega_{2D}$) in response to varying $c_S$ in Fig. 3. In the course of Raman measurements, sulfuric acid solution or DI water was intermittently added drop-wise into the optical cell to achieve gradual increase or decrease of concentration, respectively (see Fig. S5 for estimated concentration increase as a function of time). The sample was pre-undoped by submerging in DI water as described in relation to Fig. 2. Before increasing $c_S$, ($\omega_G$, $\omega_{2D}$) was located nearly on the $\varepsilon$-axis representing the degree of lattice strain, which indicates that graphene is almost charge-neutral.[15] With increasing $c_S$ to ~6 M, ($\omega_G$, $\omega_{2D}$) moved in parallel with the $n$-axis (step $i$ in Fig. 3) and reached a maximum charge density of 1.4 x $10^{13}$ cm$^{-2}$. Notably, the trajectory was quasi-linear and could be almost exactly retraced ($ii$ in Fig. 3) when $c_S$ was lowered by dilution with DI water. The average slope defined as $(\Delta\omega_{2D}/\Delta\omega_G)_n$ was 0.70 ± 0.06 from multiple samples and repeated measurements, and matched well with the value obtained from the electrically doped systems.[16] Doping and undoping could be cycled multiple times before collapsing or contamination of samples. These results validate the implicit notion[15] that charge density can be quantified with reference Raman data obtained from electrically-doped graphene systems. It is also to be noted that hysteretic difference between forward and backward trajectories is due to varying lattice strain. Thus the minor hysteresis in Fig. 3 indicates that lattice deformation was negligible during the intercalation of the dopants and its reversal unlike the first-time intercalation of water shown in Fig. 2d. We speculate that the interfacial space between graphene and substrates is filled with monolayer-thick water layer as shown in a recent study[27] and that sulfuric acid and bisulfate ions efficiently diffuse through the quasi-two-dimensional water. The interfacial molecular motion deserves further studies.

We also performed n-type doping using amine compounds both in liquid and gas phase in Fig. 3. When perforated graphene with hole density of ~5 x $10^{12}$ cm$^{-2}$ was exposed to DETA (diethylenetriamine) solution, both of $\omega_G$ and $\omega_{2D}$ decreased (yellow-filled circles) and ($\omega_G$, $\omega_{2D}$) moved along the $n$-axis reaching the $\varepsilon$-axis ($iii$ in Fig. 3), which indicated zero density or charge neutralization. As the doping proceeded further, ($\omega_G$, $\omega_{2D}$) moved horizontally to the right ($iv$ and $v$ in Fig. 3) because of accumulated n-type charge carriers. As an independent test, one pristine graphene was in-situ monitored after brief exposure to EDA (ethylenediamine) vapor. As shown by the magenta-filled circles, ($\omega_G$, $\omega_{2D}$) moved horizontally to the left finally reaching the $\varepsilon$-axis (*vi* in Fig. 3), which indicated complete desorption of EDA. While both of the n-type trajectories for DETA and EDA showed slight deviation from the electrically-induced trajectory (blue solid line), it is evident that n-type chemical doping can also be *in-situ* probed with Raman spectroscopy. However, it is to be noted that graphene membrane was not structurally stable during the treatments unlike the p-type doping. In DETA solution, graphene was highly likely to be detached or rolled up. When graphene was exposed to EDA vapor, graphene often showed change in the degree of strain. We observed that excess vapor formed micro-droplets on graphene and they are thought to exert shear force on graphene. The kink in the ($\omega_G$, $\omega_{2D}$)-trajectory for DETA (between *iv* and *v* in

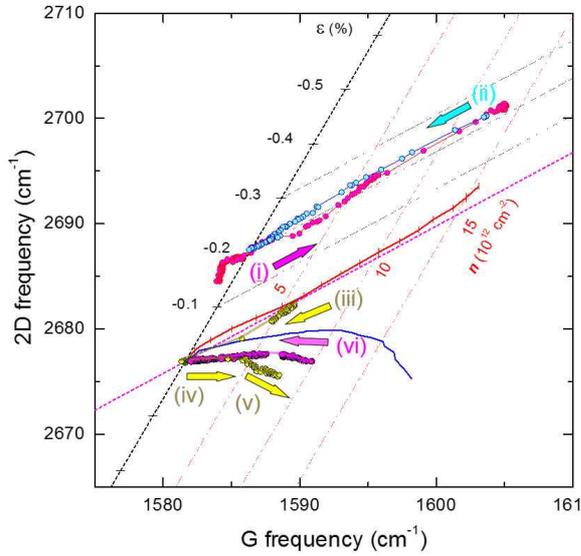

**Figure 3.** Reversible acid and base doping of nano-perforated graphene. (i & ii) To increase the degree of hole doping gradually, $H_2SO_4$ solutions of various concentration were drop-wise added during the measurement time of 80 min: 0.10 M at 5 min, 1.0 M at 20 min, 3.0 M at 35 min, 6 M at 50 min, and 9 M at 65 min. To reverse the doping, DI water was added repeatedly during the measurement time of 30 min, while the excess solution being removed by filtering paper. In order to remove the dopants generated during thermal oxidation, the sample was pre-undoped by immersing in DI water before the acid doping as shown in Fig. 2. (iii~v) p-doped perforated graphene was exposed to DETA solution and underwent charge neutralization and n-doping. (vi) Pristine graphene was in-situ monitored after brief exposure to EDA vapor. The data for iii~vi steps were displaced in parallel with the ε-axis to avoid overlap with the data for i~ii steps.

Fig. 3) can be attributed to such mechanical perturbation. By clamping graphene down on substrates, one may will be able to obtain more reliable data for the n-type trajectory.

While the calibration data in Fig. 3 was validated for chemically doped systems, they cannot be directly used in analyzing Raman spectra obtained with excitation wavelengths other than 514 nm because of the dispersive nature of $\omega_{2D}$.[23] In Fig. 4, we established analytical frameworks for popular laser lines for Raman spectroscopy. First, the origins ($\omega_G$, $\omega_{2D}$)$_o$ representing charge-neutral and unstrained graphene were empirically determined for each excitation wavelength. Since $\omega_G$ is not affected by the excitation wavelength in principle, the origins for 457 and 633 nm in Fig. 4a could be defined by displacing the origin for 514 nm[15] along the ordinate considering the dispersion in $\omega_{2D}$.[23] To validate this, freestanding samples were prepared by suspending graphene across few-micron-wide circular wells according to a previous work[15] (Fig. 4e). As shown as the origin (black circles filled in yellow) for each wavelength in Fig. 4a, $\omega_G$ averaged over the freestanding area in Fig. 4e remained virtually constant for the three wavelengths and agreed with the previous data.[15] In contrast, $\omega_{2D}$ exhibited a large variation as varying photon energy with a dispersion of $100 \pm 5$ cm$^{-1}$/eV (Fig. 4f), which agrees well with a previous work ($97 \pm 2$ cm$^{-1}$/eV) in the relevant photon energy range.[34] In particular, the newly determined origin for 514 nm (1582.2 cm$^{-1}$, 2676.6 cm$^{-1}$) was consistent with that for the previous work.[15] The origins for 457 and 633 nm were (1581.7 cm$^{-1}$, 2703.1 cm$^{-1}$) and (1581.6 cm$^{-1}$, 2629.3 cm$^{-1}$), respectively. The statistical uncertainty of the origins was smaller than spectral accuracy of the measurements (<0.5 cm$^{-1}$).

Then $\varepsilon$-axis, representing lattice strain, was defined for each excitation wavelength and validated. The peak frequencies of G and 2D are linearly dependent on strain and can be described by Gruneisen parameter ($\gamma$), $\gamma = -\frac{1}{\omega_o}\frac{\partial \omega}{\partial \varepsilon}$, where $\omega_o$ is frequency at zero strain.[14, 35] Thus $(\Delta\omega_{2D}/\Delta\omega_G)_\epsilon$ value should remain constant, which has been confirmed for native strain[15] and external biaxial strain.[36] According to Lee et al., the native strain found in exfoliated graphene (-0.2% ~ 0.4%) is better explained by uniaxial strain rather than biaxial one judging from $(\Delta\omega_{2D}/\Delta\omega_G)_\epsilon$, even though the difference is subtle for small strain values. In Fig. 4a, $\varepsilon$-axis was drawn with the reported slope[15] of $(\Delta\omega_{2D}/\Delta\omega_G)_\epsilon = 2.2 \pm 0.2$ and to pass through the origin for 514 nm. To validate this, virtually charge-neutral pristine graphene samples (one sample is shown in Fig. 4b) were prepared to obtain sets of ($\omega_G$, $\omega_{2D}$) data that contained large variation in strain values as shown in Fig. 4a. It is well known that mechanically exfoliated graphene samples show random distribution of native strain generated during the transfer.[15] Indeed, the $\omega_G$ and $\omega_{2D}$-maps obtained with 514 nm laser contained sizable frequency difference across the sample (Fig. 4c & 4d) and exhibited similar spatial variations suggesting that they are mainly affected by a common cause. When placed in Fig. 4a, the set of ($\omega_G$, $\omega_{2D}$) data for pristine sample (green circles) were found falling almost on the ε-axis (514 nm). This fact indicates that the chosen sample has sufficiently low density of extra charges and wide variation in ε, validating the ε-axis. Detailed analysis according to the previous study[15] revealed that ε ranged from -0.2% to 0.4% but *n* is less than 2 x $10^{12}$ $cm^{-2}$.

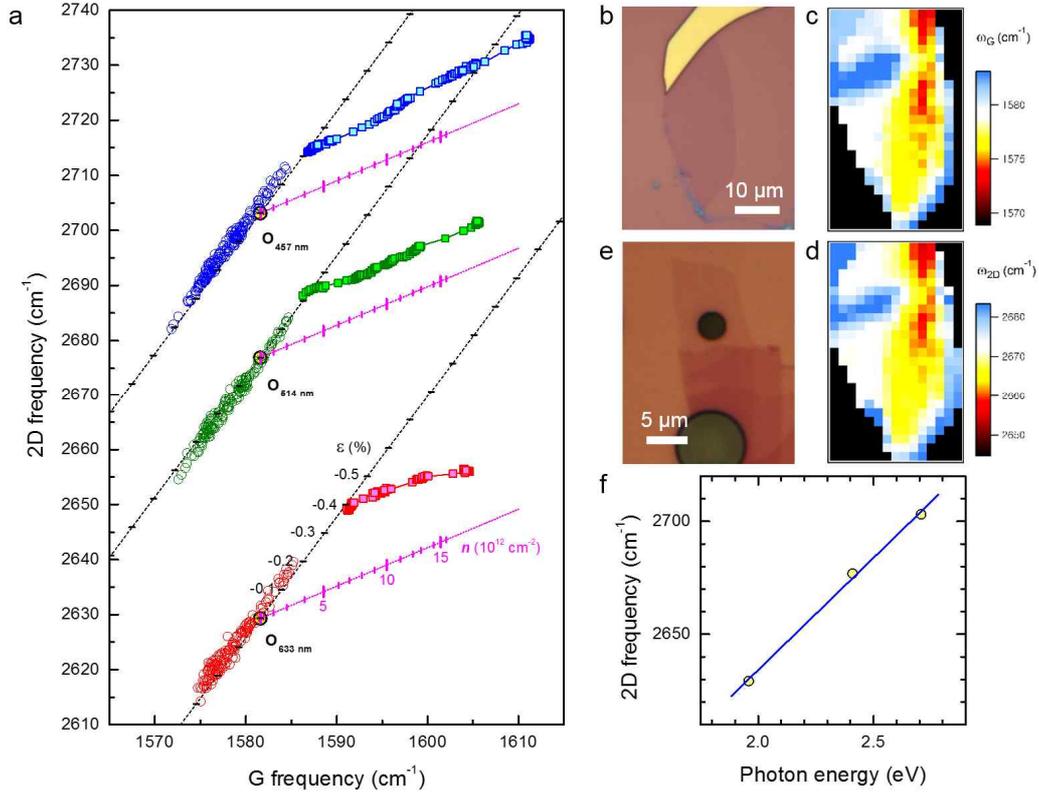

**Figure 4.** Quantification of chemical charge doping and its decoupling from lattice strain at multiple wavelengths. (a) Strain-charge density analysis for 457, 514 and 633 nm. The ($\omega_G$, $\omega_{2D}$) data in open symbols were obtained from the pristine sample in the air. Those in solid symbols were obtained for nano-perforated graphene in $H_2SO_4$ solution of varying concentration. Before the acid doping, the sample was pre-undoped after oxidation as described in Fig. 2. For the origins and ticks for the n and ε-axes, see the text. (b) Optical micrograph of pristine graphene sample exploited for the measurements. (c & d) Raman maps of $\omega_G$ (c) and $\omega_{2D}$ (d) obtained from the 1L area in (b). (e) Optical micrograph of freestanding graphene suspended across a hole of 3-micron in diameter. (f) Dispersion of $\omega_{2D}$ obtained from the freestanding sample in (e).

Figure 4a also presents two additional sets of ($\omega_G$, $\omega_{2D}$) data obtained at 457 and 633 nm from the same pristine sample. Since the G peak originates from the zone-center phonon, $\omega_G$ is not affected by the excitation wavelength. For the 2D peak, however, transverse-optical phonons of different frequency near zone-corner K are selected when excitation photon energy is varied.[37] Because of the large dispersion in $\omega_{2D}$, the slope of $\varepsilon$-axis, $(\Delta\omega_{2D}/\Delta\omega_G)_\epsilon$, is in principle dependent on excitation wavelength, even if assuming that the Gruneisen parameter of 2D modes is constant for the employed excitation wavelengths. Considering the $\omega_{2D}$-dispersion of $100 \pm 5$ cm$^{-1}$/eV (Fig. 4f), $\omega_{2D}$ at zero strain varies 75 cm$^{-1}$ between 457 and 633 nm. This translates into a negligible change in $(\Delta\omega_{2D}/\Delta\omega_G)_\epsilon$ (< 3% between 457 and 633 nm) compared to the experimental uncertainty of 10%.[15] Thus, the $\varepsilon$-axes for all three wavelengths in Fig. 4a were set to have the same slope of 2.2 that was derived for 514 nm.[15] The ticks on the $\varepsilon$-axes for 457 and 633 nm were generated by displacing those for 514 nm by the displacement in their origins. Indeed, all the data points in Fig. 4a were aligned along the $\varepsilon$-axis defined for each excitation wavelength. From linear regression analyses, slopes of 2.22, 2.31 and 2.14 were obtained for the three data sets in the order of increasing wavelength.

Finally, *n*-axes in Fig. 4a were also validated for other wavelengths by H$_2$SO$_4$-doping. For G and 2D peaks, their frequency dependence on charge density has been well studied theoretically[38] and experimentally.[8, 16] Both approaches reached a consensus that the hole-induced fractional frequency change, $(\Delta\omega_{2D}/\Delta\omega_G)_n$, is nearly constant as $0.70 \pm 0.06$ in a wide range of hole density (0 ~ 1.5x10$^{13}$ cm$^{-2}$).[15] However, it has not been clear that the same value can be used for excitation wavelengths other than 514 nm. While $\omega_G$ is not affected as discussed earlier, the dependence of $\omega_{2D}$ on *n* may be influenced by photon energy, in other words, $\omega_{2D}$ of different momentum may change differently upon injection of extra charges. Unfortunately, however, no experimental and theoretical investigations were available. As a first-order approximation, one may argue that $\Delta\omega_{2D}/\Delta n \propto \omega_{2D}^{n=0}$. Since $\omega_{2D}^{n=0}$ varies 75 cm$^{-1}$ between 457 and 633 nm, $(\Delta\omega_{2D}/\Delta\omega_G)_n$ will vary by less than 3%. This consideration indicates that $(\Delta\omega_{2D}/\Delta\omega_G)_n$ of $0.70 \pm 0.06$ will be valid for all visible excitation wavelengths within the experimental error. Figure 4a presents *n*-axes for 457 and 633 nm that are drawn with a slope of 0.70 from each origin. The ticks on the *n*-axes for 457 and 633 nm were also generated by displacing those for 514 nm by the displacement in their origins. For an experimental confirmation, the pristine graphene presented in Fig. 4a was nano-perforated by oxidation and subjected to cycles of undoping and doping with in-situ Raman measurements (see Fig. S6 for Raman spectra at varying charge density obtained for three wavelengths). The trajectories of ($\omega_G$, $\omega_{2D}$) for the three wavelengths were parallel to each other and essentially to the *n*-axes. This observation assures that quantification of charge density can also be made at excitation wavelengths other than 514 nm. The slight nonlinearity for 457 and 633 nm was attributed to the fact that the focal spot slightly drifted within the sample area with spatial inhomogeneity.

The Raman spectroscopic calibration data provided in this work were verified against native strain and chemical doping by sulfuric acid, and theoretical prediction. They can be readily applied in quantifying strain and charge density in graphene using the three wavelengths and possibly arbitrary wavelength when the dispersion in $\omega_{2D}$ is corrected. Moreover, the graphene membrane system presented in this work will be useful in investigating various chemical interaction of graphene including charge transfer reactions as demonstrated in this work. Since the membrane is made of high quality graphene mechanically exfoliated from bulk crystals, its Raman spectra are highly reproducible and of narrow linewidth, which leads to high precision in various material properties obtained by Raman spectroscopy. In addition, graphene membranes are precisely perforated with dense arrays of nanopores that allow efficient intercalation of molecular species through graphene-substrate interfaces. Despite the interfacial diffusion, the membrane system was robust for multiple cycles of doping and undoping in aqueous solutions. Mechanistic details of acid doping and molecular motions at the interface are still unclear and deserve future study.

## 4. Conclusions.

We created a simple but reliable nano-perforated graphene membrane system, and studied how G and 2D Raman peaks shift in response to lattice strain and chemical charge doping by sulfuric acid solution. Raman spectroscopic metrology for strain and charge density was separately established and confirmed for three popular excitation wavelengths of 457, 514 and 633 nm because of the dispersing 2D peak. Freestanding graphene was prepared and served as an unperturbed graphene. Randomly distributed native strain in pristine samples was exploited in deriving strain-dependent fractional frequency change in G and 2D peaks, $(\Delta\omega_{2D}/\Delta\omega_G)_\epsilon$ = 2.2 ± 0.2. Charge density-dependent fractional frequency change in G and 2D peaks, $(\Delta\omega_{2D}/\Delta\omega_G)_n$ = 0.70 ± 0.06, was determined by in-situ doping and undoping of graphene by sulfuric acid. Despite the dispersion in $\omega_{2D}$, the slopes were identical for all the employed wavelengths within the experimental uncertainty.

## 5. Conflict of interest

The authors declare no conflict of interest.

**ACKNOWLEDGMENTS**

This work was supported by the National Research Foundation of Korea (NRF-2015R1A2A1A15052078 and NRF-2016R1A2B3010390) and POSCO Green Science Project.

## Appendix A. Supplementary data

Supplementary data associated with this article can be found, in the online version.